\documentclass[12pt]{article}
\usepackage{feynmp,slashed}
\usepackage{amssymb,graphicx}
\usepackage{graphics}
\usepackage{amsmath,amsfonts}

\def\e{{\rm e}}

\def\d{\partial}
\def\l{\left(}
\def\r{\right)}

\def\g{\gamma}
\def\vk{\varkappa}
\newcommand{\be}{\begin{equation}}
\newcommand{\ee}{\end{equation}}
\newcommand{\bea}{\begin{eqnarray}}
\newcommand{\eea}{\end{eqnarray}}
\newcommand{\bg}{\begin{gather}}
\newcommand{\eg}{\end{gather}}
\newcommand{\bseq}{\begin{subequations}}
\newcommand{\eseq}{\end{subequations}}

\newcommand\m{\mu}
\newcommand\D{\Delta}
\newcommand\n{\nu}
\renewcommand\l{\lambda}
\renewcommand\r{\rho}

\title{
\sc{\huge On calculation of cross sections in Lorentz violating theories}
}
\author{
  Grigory Rubtsov$^{a,b}$, Petr Satunin$^a$ and Sergey Sibiryakov$^{a,b}$
\vspace{.2cm}\\
\normalsize\llap{$^a$}\it Institute for Nuclear Research of the
Russian Academy of Sciences, \\ 
      \normalsize \it  60th October Anniversary Prospect, 7a, 117312
      Moscow, Russia\\
\normalsize\llap{$^b$}\it Faculty of Physics, Moscow State University,
\normalsize \it Vorobjevy Gory, 119991 Moscow, Russia
} 

\begin{document}
\maketitle

\begin{abstract}
We develop a systematic approach to the calculation of scattering
cross sections in theories 
with violation of the Lorentz invariance taking into account 
the whole information about the theory
Lagrangian. 
As an illustration we derive the Feynman
rules and formulas for sums over polarizations in spinor
electrodynamics with Lorentz violating operators of dimensions four
and six. These rules are applied to compute the probabilities of 
several astrophysically relevant processes.
We calculate the rates of photon decay and vacuum Cherenkov
radiation along with the cross sections of electron-positron pair
production on background radiation and in the Coulomb
field. The latter process is essential for detection of
photon-induced air showers in the atmosphere. 
\end{abstract}

\section{Introduction}
\label{sec:intro}

A number of approaches to quantum gravity suggest that Lorentz
invariance (LI) may be not exact and breaks down at high energies, see
\cite{Mattingly:2005re} and references therein (see also
\cite{Horava:2009uw}). 
Within
the effective field theory approach the deviations from LI are described by
higher dimension operators suppressed by the putative scale $M$ of quantum
gravity. This scale is supposed to coincide with the Planck mass,
$M_{P}\approx 10^{19}$~GeV, in most approaches, but can 
lie a few orders of magnitude below in certain scenarios
\cite{Blas:2009qj}.  
Important
constraints on the high-energy violation of LI (Lorentz violation (LV)
for short) have been obtained from considerations of various
astrophysical phenomena
\cite{Jacobson:2005bg,Liberati:2009pf}. Indeed, in the astrophysical
processes the elementary particles often reach energies that vastly
exceed those attained in the accelerator experiments. Therefore these
processes provide a unique probe of the particle dynamics at very high
energy. 
The
extreme energies ever observed 
are reached by ultra-high-energy cosmic rays (UHECR).
The power of UHECR physics in constraining Planck--suppressed LV 
has been extensively discussed in the literature
\cite{Galaverni:2007tq,Galaverni:2008yj,Maccione:2008iw,Maccione:2009ju,
Mattingly:2009jf,Saveliev:2011vw}. 

Most of these studies concentrate on the kinematical effects of LV:
appearance
of new reactions that are kinematically forbidden in the LI
case and the shift of energy thresholds of the known
processes. Clearly, more information can be gained by considering the
dynamical consequences of LV, i.e. the effect of LV on various
reaction rates. This is particularly important in the case of reactions 
that do not
possess a threshold such as photon splitting into several photons 
\cite{Gelmini:2005gy} and
neutrino splitting \cite{Mattingly:2009jf} 
(see also \cite{Maccione:2011fr}). Another example is
provided by the Bethe--Heitler process --- 
production of electron-positron pairs by a photon 
in the Coulomb field of a nucleus --- that plays the key role 
in 
the detection and identification of UHECR photons through their
interaction with the Earth atmosphere \cite{Risse:2007sd}.
The dynamical analysis requires developing a technique for evaluation
of the Feynman diagrams 
in LV theories with higher order operators. 
So far, no systematic treatment of this issue has been performed.
The aim of the present paper is to fill this gap.  

LV affects the rate of a given process in three ways: {\it (i)}
through modification of the phase space integrals; 
{\it (ii)} different wave-functions of the external states;
{\it (iii)} changes in the vertices and propagators. 
The effects {\it (ii)} and {\it (iii)} lead to 
modification of the expression for the matrix element of the process
compared to the standard LI case. We are going to see
that all three effects are of the same order and must be taken into
account simultaneously to obtain the correct result. 
As an illustration of our
technique we will
present calculations of the rates of several astrophysically relevant processes
in spinor quantum electrodynamics (QED) with LV operators of dimension
four and six. The results will be compared with the estimates based on
the kinematical considerations.

Let us mention several works relevant for our study. The modifications
of the phase space
and the external-states wavefunctions were discussed in
\cite{Jacobson:2005bg} in the context of QED with dimension-five LV
terms. Spin sums over external states for the model similar to the one
considered in this paper were derived in \cite{Anselmi:2011ae}. 
Ref.~\cite{Klinkhamer:2008ky} considers QED with LV restricted to 
dimension four operators and calculates the rates of  
the vacuum Cherenkov radiation and photon decay into electron-positron
pair in this theory. The necessity to take all effects {\it (i)}
-- {\it (iii)} into account was recently stressed in
\cite{Bezrukov:2011qn,Carmona:2012tp} in the context of theories with
LV in the
neutrino sector. 

The paper is organized as follows. In Sec.~\ref{sec:rules} we
introduce the model, derive the formulas for the spin sums over
external states and the Feynman rules. In Sec.~\ref{sec:rates} we
apply these rules to calculate the rates of several processes:
photon decay, vacuum Cherenkov radiation, 
pair production in two-photon collision and pair-production by a
photon  
in the Coulomb
field. 
Sec.~\ref{sec:discussion} is devoted to the discussion of our results.

\section{The model: spin sums and Feynman rules}
\label{sec:rules}
We are going to consider QED with LV operators of dimension up to 6
assuming that the gauge invariance is preserved. 
To simplify the
analysis we impose several additional restrictions:
\begin{itemize}
\item[(a)] The theory is required to be invariant under rotations in
  the three-dimensional space.
\item[(b)] We impose the CPT and P invariance. The CPT symmetry is
  physically essential as it forbids dangerous dimension 3 operators
  that would lead to an unacceptably large LV at low energies
  \cite{Kostelecky:2008ts}. On the 
  other hand, the requirement of spatial parity is purely technical
  and is invoked just to further reduce the number of LV structures. It can be
  easily dropped in a more general treatment.
\item[(c)] We include in the Lagrangian only operators that cannot be
  removed by a field and/or coordinate redefinition. 
\item[(d)] According to the general logic of the effective field
  theory, the higher-dimensional operators are equivalent if
  they coincide on the equations of motion obtained from the
  lower-dimensional part of the Lagrangian. We consider only
  operators that are different with respect to this identification. 
\item[(e)] Above the electroweak scale $M_{EW}$ the model must be
  embedded into the full Standard Model. The chiral structure of the
  latter forbids CPT-even LV operators of dimension~5~
  \cite{Bolokhov:2007yc}. This means that even though such terms
can be
generated below $M_{EW}$, the dimensionless coefficients in front of
them will be suppressed by the ratio $M_{EW}/M$ 
making their contribution negligible. We do not consider 
dimension 5 operators below.
\item[(f)] We include only 
operators that contain parts quadratic in the fields meaning that
they contribute into the free-particle Lagrangian. The rationale behind 
this requirement is
purely pragmatic: to have a minimal framework where LV affects both
kinematics and dynamics of the theory.   
\end{itemize}
The above conditions lead 
to the following Lagrangian, cf. \cite{Mattingly:2008pw},
\be
\label{lagr}
{\cal L}=i\bar\psi\g^\m D_\m\psi-m\bar\psi\psi-\frac{1}{4}F_{\m\n}F^{\m\n}
+i\varkappa\bar\psi\g^iD_i\psi
+\frac{ig}{M^2}D_j\bar\psi\g^iD_iD_j\psi
+\frac{\xi}{4M^2}F_{kj}\d_i^2F^{kj}\;,
\ee
where
\[
D_\m\psi=(\d_\mu+ieA_\m)\psi\;.
\]
is the standard covariant derivative.
The Greek indices $\mu,\nu,\ldots$ run from $0$ to $3$ and are
raised/lowered using the Minkowski metric with the signature $(+,-,-,-)$.
The Latin indices $i,j,\ldots$ are purely space-like and 
take the values $1,2,3$. Notice that we distinguish the lower and upper
space indices: $\g^i=-\g_i$, etc. Summation
over repeated indices is understood. 
The last three terms in (\ref{lagr}) describe LV. The first of them
has dimension 4 while the other two have dimension 6 and
are
suppressed by the LV scale $M$. The
dimensionless parameters $\vk$, $g$, $\xi$ characterize the strength of LV
(one of the parameters $g$, $\xi$ can be absorbed into
redefinition of the scale $M$ but we prefer to keep them
explicitly). From the viewpoint of the effective theory,
the dimension 4 operator
should be treated as the
leading LV term.
However, 
the corresponding coefficient    
$\varkappa$ is experimentally constrained to be extremely small\footnote{We
do not address the naturalness issues related to the
smallness of $\vk$, see e.g. the discussion in \cite{Liberati:2009pf}.},
$|\vk|<10^{-15}$ \cite{Kostelecky:2008ts}, implying that for
astrophysically relevant processes the effects of the dimension 6
operators can be comparable or even dominant\footnote{For example, 
for UHECR 
energies $E\sim 10^{19}$~eV and $M=10^{16}$~GeV we obtain 
$(E/M)^2\sim 10^{-12}\gg \vk$.}.

Consider the free-particle states. The LV terms
modify the dispersion relations
for photons
and electrons / positrons,
\begin{align}
\label{Eg}
&E_\g^2=k^2+\frac{\xi k^4}{M^2}\;,\\
\label{Ee}
&E_e^2=m^2
+p^2\bigg(1+\vk+\frac{gp^2}{M^2}\bigg)^2\approx
m^2+p^2(1+2\vk)+\frac{2gp^4}{M^2}\;.
\end{align}
We observe two types of modifications. First, at $p\ll M$ the velocity of
electrons is different from that of photons. Second, at large energies
the dispersion relations receive contributions that are quartic 
in the particle momenta.
Note that with our conventions the velocity of the low-energy photons
is equal to 1.

Our immediate task is determination of the particles' wavefunctions
together with 
the sums over polarizations needed for the calculation of the
inclusive cross sections. 
First, consider photon.
In the frame
where its three-momentum $k^i$ is directed along the third axis
the two polarization vectors are
\[
\varepsilon_\m^{(1)}=(0,1,0,0)~,~~~~\varepsilon_\m^{(2)}=(0,0,1,0)\;.
\]
This gives the standard formula
\[
\sum_{a=1,2} \varepsilon^{(a)}_\m\varepsilon^{(a)}_\n
=diag(0,1,1,0)\;.
\] 
We want to cast it into the form that preserves three-dimensional
rotational invariance. To do this we use the gauge invariance that
ensures that photon couples to a conserved current. Thus the Ward
identities remain valid and 
one can add
to the
above expression any combination of the form 
$n_\m k_\n+k_\m n_\n$ with an arbitrary
vector $n_\m$ without affecting the matrix element. 
A proper choice of $n_\m$ allows to cast the above sum into the form,
\be
\label{Ximunu}
\sum_{a=1,2} \varepsilon^{(a)}_\m\varepsilon^{(a)}_\n\simeq
 diag(-E_\g^2/k^2,1,1,1)\;.
\ee
This reduces to the standard expression 
$$\sum_{a=1,2} \varepsilon^{(a)}_\m\varepsilon^{(a)}_\n\simeq -\eta_{\m\n}$$
in the case of the LI dispersion relation $E_\g=k$.

Let us turn now to the spinor wave-functions. Decomposing the spinor
field into positive and negative frequency components,
\[
\psi(x,t)=\int \frac{d^3p}{(2\pi)^3\sqrt{2E}}
\Big(\e^{-iEt+ipx}u^s(p)a_s(p)+\e^{iEt-ipx}v^s(p)b^+_s(p)\Big)\;,
\]
we find the solutions of the modified Dirac equation:
\[
u^s(p)=\begin{pmatrix}
\sqrt{E-(\sigma^ip^i)\bigg(1+\vk+\frac{gp^2}{M^2}\bigg)}\,\chi^s\\
\sqrt{E+(\sigma^ip^i)\bigg(1+\vk+\frac{gp^2}{M^2}\bigg)}\,\chi^s
\end{pmatrix}~,~~~
v^s(p)=\begin{pmatrix}
\sqrt{E-(\sigma^ip^i)\bigg(1+\vk+\frac{gp^2}{M^2}\bigg)}\,\zeta^s\\
-\sqrt{E+(\sigma^ip^i)\bigg(1+\vk+\frac{gp^2}{M^2}\bigg)}\,\zeta^s
\end{pmatrix}
\]
where $\sigma^i$ are the Pauli matrices and 
$\chi^s$ and $\zeta^s$, $s=1,2$ are two-component basis
spinors. Choosing the latter to be orthogonal with unit norm we obtain 
the usual normalization conditions,
\[
\big(u^r(p)\big)^\dagger u^s(p)=\big(v^r(p)\big)^\dagger v^s(p)=2E\delta^{rs}~,~~~~
\big(u^r(p)\big)^\dagger v^s(-p)=0\;.
\] 
This leads to the spin sums:
\bseq
\label{UV}
\begin{align}
\label{U}
\sum_{s=1,2}u^s(p)\bar u^s(p)
=\g^0 E-\g^i
p^i\bigg(1+\vk+\frac{gp^2}{M^2}\bigg)+m\;,\\
\label{V}
\sum_{s=1,2}v^s(p)\bar v^s(p)
=\g^0 E-\g^i
p^i\bigg(1+\vk+\frac{gp^2}{M^2}\bigg)-m\;.
\end{align}
\eseq
One can check 
using these expressions 
and the standard creation--annihilation algebra
\[
\{a_s(p),a_r^+(q)\}=\{b_s(p),b_r^+(q)\}=(2\pi)^3\delta_{sr}\delta(p-q)\;
\]
that the spinor operators
satisfy the canonical commutation relations
\[
\{\psi(x,t), \psi^\dagger(y,t)\}=\delta(x-y)\;.
\]
It is convenient to introduce the notations
\be
\label{tildep}
\tilde p^0=E~,~~~~
\tilde p^i=p^i\bigg(1+\vk+\frac{gp^2}{M^2}\bigg)\;
\ee
that allows to write (\ref{UV}) in a more compact form
\[
\sum_{s=1,2}u^s(p)\bar u^s(p)
=\g^\m\tilde p_\m+m\;,~~~~~
\sum_{s=1,2}v^s(p)\bar v^s(p)
=\g^\m\tilde p_\m-m\;.
\] 
The formulas (\ref{Ximunu}), (\ref{UV}) were previously derived 
in \cite{Anselmi:2011ae}.

To compute the cross sections we need 
the Feynman rules for the theory (\ref{lagr}). 
In terms of the vector $\tilde p^\m$ the fermion propagator is given
by the
formally standard expression,
\[
\begin{fmffile}{ferm}
\parbox{50pt}{\begin{fmfgraph*}(50,9)
\fmfpen{thick}
\fmfleft{i} 
\fmfright{o}
\fmf{fermion,label=$p$,label.side=right}{i,o}
\end{fmfgraph*}}~~~~
=~~\frac{i(\g^\m\tilde p_\m+m)}{\tilde p_\m\tilde p^\m-m^2+i\epsilon}\;.
\end{fmffile}
\]
To obtain the photon propagator\footnote{
We will not use the photon propagator in the rest of the
paper and present it here only for the sake of completeness.} 
we have to choose the gauge. A 
convenient gauge fixing term, that eliminates non-diagonal
contributions in the photon Lagrangian, is
\be
\label{LGF}
{\cal L}_{GF}=-\frac{1}{2}
\bigg(\d_0A_0-\bigg(1-\frac{\xi}{M^2}\D\bigg)\d_i A_i\bigg)
\bigg[1-\frac{\xi}{M^2}\D\bigg]^{-1}
\bigg(\d_0A_0-\bigg(1-\frac{\xi}{M^2}\D\bigg)\d_i A_i\bigg)\;,
\ee 
where $\D\equiv \d_i\d_i$ is the spatial Laplacian. The resulting
propagator has the form,
\[
\begin{fmffile}{phot}
\parbox{50pt}{\begin{fmfgraph*}(50,9)
\fmfpen{thick}
\fmfleft{i} 
\fmfright{o}
\fmf{photon,label=$k$,label.side=right}{i,o}
\end{fmfgraph*}}~~~~
=~~i\bigg[E^2-k^2\bigg(1+\frac{\xi
  k^2}{M^2}\bigg)+i\epsilon\bigg]^{-1}
diag\bigg(-\bigg(1+\frac{\xi k^2}{M^2}\bigg),1,1,1\bigg)\;,
\end{fmffile}
\]
Note that the gauge-fixing 
term (\ref{LGF}) is non-local in space. However, one does not expect
any problems related to that, at least within the perturbation
theory. As an alternative, one can consider local gauge fixing 
and work with an off-diagonal photon propagator.

It remains to obtain the expressions for the interaction vertices. The
fourth and fifth terms in (\ref{lagr}) modify the 
photon--fermion interaction that now takes the form
\be
\label{1phot}
\begin{fmffile}{1phot}
{\cal V}_{1\g}^{\m}\equiv
\parbox{80pt}{\begin{fmfgraph*}(80,80)
\fmfpen{thick}
\fmfleft{i}
\fmfrightn{o}{2}
\fmf{photon,tension=1.4}{i,v}
\fmf{fermion,label=$p_2$,label.side=right}{o1,v}
\fmf{fermion,label=$p_1$,label.side=right}{v,o2}
\end{fmfgraph*}}
=-ie\g^\m-ie\delta^\m_i\bigg[\vk\g^i+
\frac{g}{M^2}\big(p_1^i(p_1\cdot\g)+p_2^i(p_2\cdot\g)
-(p_1\cdot p_2)\g^i\big)\bigg]\;,
\end{fmffile}
\ee
where the two fermion momenta 
$p_1$, $p_2$ are assumed to flow {\em out} of the
vertex and
dot stands for the scalar product of three-dimensional
vectors,
$(p_1\cdot\g)=p^i_1\g^i$, etc. 
Besides, the fifth term in (\ref{lagr}) introduces new vertices
involving two and three photons,
\begin{align}
\label{2phot}
&\begin{fmffile}{2phot}
{\cal V}_{2\g}^{\m\n}\equiv
\parbox{80pt}{\begin{fmfgraph*}(80,70)
\fmfpen{thick}
\fmfleftn{i}{2}
\fmfrightn{o}{2}
\fmf{photon}{i1,v}
\fmf{photon}{i2,v}
\fmf{fermion,label=$p_2$,label.side=right}{o1,v}
\fmf{fermion,label=$p_1$,label.side=right}{v,o2}
\end{fmfgraph*}}
=\frac{ige^2}{M^2}\Big[\g^i(p_2-p_1)^j+\g^j(p_2-p_1)^i
+\delta^{ij}\big((p_2-p_1)\cdot\g\big)\Big]\delta^\m_i\delta^\n_j\;,
\end{fmffile}\\
\notag\\
\label{3phot}
&\begin{fmffile}{3phot}
{\cal V}_{3\g}^{\m\n\l}\equiv
\parbox{80pt}{\begin{fmfgraph*}(80,70)
\fmfpen{thick}
\fmfleftn{i}{3}
\fmfrightn{o}{2}
\fmf{photon}{i1,v}
\fmf{photon,tension=0}{i2,v}
\fmf{photon}{i3,v}
\fmf{fermion,label=$p_2$,label.side=right}{o1,v}
\fmf{fermion,label=$p_1$,label.side=right}{v,o2}
\end{fmfgraph*}}
=-\frac{2ige^3}{M^2}\Big[\delta_i^\m\delta_j^\n\delta_j^\l
+\delta_i^\n\delta_j^\m\delta_j^\l+\delta_i^\l\delta_j^\m\delta_j^\n
\Big]\g^i\;,
\end{fmffile}
\end{align}
where the momenta $p_1$ and $p_2$ are again flowing {\em out} of the
vertex. Note that the 2-photon interaction (\ref{2phot}) 
is antisymmetric under the exchange of the electron and
positron momenta. It is worth stressing that the modification of the
1-photon interaction and the presence of the multi-photon 
vertices (\ref{2phot}),
(\ref{3phot}) are required by the gauge invariance and ensure that the
Ward
identities are satisfied.  

We are now going to apply the above rules to compute the rates of several
reactions.

\section{Processes and rates}
\label{sec:rates}

We start with the elementary processes of photon decay and vacuum
Cherenkov radiation. From the viewpoint of the astrophysical
applications the exact expressions for the rates in these cases are
unnecessary. Indeed, these processes are extremely fast, once
kinematically allowed, and for all practical purposes may be
considered as happening instantaneously. However, the study of these
simple reactions is instructive and sets the stage for the more
involved calculations in the next subsections. 

\subsection{Photon decay}
\label{subsec:decay}

The photon decay
\be
\label{gdecay}
\gamma\to e^+e^-
\ee
can become allowed in the presence of LV above certain
threshold, see e.g. Ref.~\cite{Galaverni:2008yj} for the discussion of
the kinematics of this reaction. We are interested in computing the
rate of the process well above the threshold. In this regime the masses
of the electron and positron can be neglected which 
considerably simplifies the calculation. The matrix element has
the form
\[
{\cal M}=\bar u(p_1)\,{\cal V}_{1\g}^\m\, v(p_2)\,\varepsilon_\m\;,
\]
with the vertex given by Eq.~(\ref{1phot}). The inclusive rate is
obtained by taking the 
square of this expression, summing over the spins of the final states and
averaging over the photon polarizations. Using (\ref{Ximunu}),
(\ref{UV}), where we neglect the fermion mass, and keeping only up to
linear terms in the LV parameters we obtain,
\be
\label{Mxig1}
\begin{split}
\overline{|{\cal M}|^2}=&4e^2(1+3\vk)(E_1E_2-(p_1\cdot p_2))\\
&-\frac{2e^2\xi}{M^2}k^2(E_1E_2+(p_1\cdot p_2))
+\frac{4e^2g}{M^2}(p_1 \cdot p_2)^2
+\frac{4e^2g}{M^2}E_1E_2(p_1^2+p_2^2-3(p_1\cdot p_2))\;.
\end{split}
\ee 
The first line here is the standard matrix element of the Lorentz
invariant QED multiplied by an overall $\vk$-dependent factor. We will
see shortly that in the leading approximation the dependence of such 
overall factors on the LV parameters can be safely neglected. 
On the other hand, the second line represents the genuine LV
correction 
to the matrix element. 

The decay width is given by the textbook formula,
\[
\Gamma=\frac{1}{2E_\g}\int \frac{d^3p_1}{(2\pi)^3 2E_1}
\frac{d^3p_2}{(2\pi)^3 2E_2} (2\pi)^4\delta(E_\g-E_1-E_2)\,
\delta^{(3)}(k-p_1-p_2)\,\overline{|{\cal M}|^2}\;,
\] 
that remains valid in
the presence of LV (its derivation does not make any use of LI).
One chooses the photon momentum to be directed along the first axis
and parameterizes the momenta of the fermions as follows,
\[
p_1^i=\big(k(1+x)/2,\,p_\bot,0\big)~,~~~
p_2^i=\big(k(1-x)/2,-p_\bot,0\big)\;.
\] 
Clearly, up to rotations, this is the most general parameterization
satisfying the momentum conservation. The conservation of energy in
the zeroth order in the LV parameters requires $x$ to lie in the range
$-1\leq x\leq 1$. In the ultrarelativistic regime, that is of interest
to us, we have $p_\bot\ll k$. This allows to expand
\[
E_\g=k+\frac{\xi k^3}{2M^2}\;,~~~~
E_{1,2}=\frac{k}{2}(1+\vk)(1\pm x)+\frac{gk^3}{8M^2}(1\pm
x)^3+\frac{p_\bot^2}{k(1\pm x)}\;,
\]
which yields
\be
\label{gdec}
\begin{split}
\Gamma=\frac{1}{2k}\int
\frac{dxdp_\bot^2}{8\pi k(1-x^2)}\,
\delta\bigg(\omega_{LV}(x)
-\frac{2p_\bot^2}{k(1-x^2)}\bigg) \,\overline{|{\cal M}|^2}\;,
\end{split}
\ee 
where we have introduced the notation
\be
\label{omegaLV}
\omega_{LV}(x)=-\vk k+\frac{\xi k^3}{2M^2}-\frac{gk^3}{4M^2}(1+3x^2)\;.
\ee
It is instructive to compute the
contributions of the first (Lorentz invariant) and second 
(Lorentz violating) lines of (\ref{Mxig1}) 
into the decay rate separately. For the 
contribution of the LI part of the matrix element we have,
\be
\label{gdecstand}
\begin{split}
\Gamma_1&=\frac{\alpha(1+3\vk)}{k}\int\!\!
\frac{dxdp_\bot^2}{k(1-x^2)}
\delta\bigg(\omega_{LV}(x)
-\frac{2p_\bot^2}{k(1-x^2)}\bigg)
\bigg[\frac{\vk k^2}{2}(1-x^2)
+\frac{gk^4}{8M^2}(1-x^4)+\frac{2p_\bot^2}{1-x^2}\bigg]\\
&=\alpha\, k
\int dx\bigg[-\frac{\vk}{4}(1+x^2)+\frac{k^2}{M^2}\bigg(\frac{\xi}{4}
-\frac{g}{16}-\frac{3gx^2}{8}-\frac{gx^4}{16}\bigg)\bigg]\;,
\end{split}
\ee 
where $\alpha=e^2/4\pi$ is the fine structure constant and passing to
the second line we performed the integration of $p_\bot$. 
We see that the rate is suppressed by the parameters describing LV. 
Notice, however,
that the
suppression does not come from the phase space integration, that reduces
to the integral over $x$. Instead, the suppression is due to the smallness
of the LI matrix element for the nearly collinear kinematics
realized in the decay. For the contribution of the terms in the second
line of (\ref{Mxig1}) we obtain,
\[
\Gamma_{2}=\alpha\, k
\int dx \,\frac{k^2}{M^2}\bigg(-\frac{\xi}{8}+\frac{\xi x^2}{8}
+\frac{g x^2}{8}-\frac{g x^4}{8}\bigg)\;.
\] 
One observes that it is of the same order as
(\ref{gdecstand}). Combining the two contributions we obtain the total
decay rate,
\be
\label{gdectot}
\Gamma_{\g\to e^+e^-}=\frac{\alpha}{4}
\int dx\, (1+x^2)\,\omega_{LV}(x)\;.
\ee

The domain of integration in $x$ is determined by the condition
that $p_\bot^2$ found from the $\delta$-function in
(\ref{gdec}) is positive. This implies
\[
\omega_{LV}(x)\geq 0\;.
\]
One distinguishes several cases:\\
{\it (a)} $0\leq 2g\leq\xi_*$ or 
$g\leq 0$, $g/2\leq\xi_*$, where $\xi_*=\xi-2M^2\vk/k^2$. 
In this case the integration is over the whole range $-1\leq
x\leq 1$ yielding
\[
\Gamma_{\g\to e^+e^-}=\alpha\,k\bigg[-\frac{2\vk}{3}+\frac{k^2}{M^2}
\bigg(\frac{\xi}{3}-\frac{11g}{30}\bigg)\bigg]\;.
\]
For $\xi=g=0$, $\vk<0$ this coincides with the result 
of\footnote{In \cite{Klinkhamer:2008ky} 
the rate is calculated in the model where the
  electron/positron have unit velocities while the photon velocity
  differs from one. This is related to our setup by a rescaling of
  the space coordinates and therefore is physically equivalent.} 
\cite{Klinkhamer:2008ky}.\\
{\it (b)} $0<g$, $g/2<\xi_*<2g$. The integration range is 
\[
-\frac{1}{\sqrt 3}\sqrt{\frac{2\xi_*}{g}-1}\leq x \leq
\frac{1}{\sqrt 3}\sqrt{\frac{2\xi_*}{g}-1}
\]
which gives
\[
\Gamma_{\g\to e^+e^-}=\frac{\alpha k^3}{90\sqrt{3}M^2}
\sqrt{\frac{2\xi_*}{g}-1}\;\bigg(13\xi_*-7g+\frac{2\xi_*^2}{g}\bigg)\;.
\]
Note that in this case the fraction of the total energy carried by
each fermion is bounded from below by a strictly positive number.\\
{\it (c)} $2g<\xi_*<g/2<0$. 
The integration is performed over two disjoint regions
\[
-1\leq x\leq-\frac{1}{\sqrt 3}\sqrt{\frac{2\xi_*}{g}-1}~~~~~
\mathrm{and}~~~~~
\frac{1}{\sqrt 3}\sqrt{\frac{2\xi_*}{g}-1}\leq x\leq1\;
\]
implying that the electron and positron momenta are necessarily
different. This corresponds to the regime of the 
the so-called ``asymmetric threshold'' \cite{Mattingly:2002ba}. In
this case we have
\[
\Gamma_{\g\to e^+e^-}=\frac{\alpha k^3}{3M^2}
\bigg[\bigg(\xi_*-\frac{11}{10}g\bigg)
-\frac{1}{30\sqrt{3}}\sqrt{\frac{2\xi_*}{g}-1}
\bigg(13\xi_*-7g+\frac{2\xi_*^2}{g}\bigg)\bigg]\;.
\]

\subsection{Cherenkov radiation}
\label{subsec:cherenkov}

Another elementary process that can become allowed in the presence of
LV is the vacuum Cherenkov radiation --- 
spontaneous emission of a photon by a fast moving electron, 
\[
\e^-\to\gamma e^-\;.
\]
This is the cross-channel of the reaction (\ref{gdecay}). As before,
we consider the rate well above the threshold and thus neglect the 
electron mass. Thus the
corresponding matrix element is obtained from (\ref{Mxig1}) by 
replacing the positron energy and momentum with the (minus) energy and 
momentum of the incoming electron,
\[
E_2\mapsto -E~,~~~~p_2^i\mapsto -p^i\;,
\]
and by flipping the overall sign. This yields
\be
\label{Mxcher}
\begin{split}
\overline{|{\cal M}|^2}=&4e^2(1+3\vk)(E E'-(p\cdot p'))\\
&-\frac{2e^2\xi}{M^2}k^2(E E'+(p\cdot p'))
-\frac{4e^2g}{M^2}(p\cdot p')^2
+\frac{4e^2g}{M^2}E E'(p^2+{p'}^2+3(p\cdot p'))\;,
\end{split}
\ee
where $E'$, ${p'}^i$ are the energy and
momentum of the outgoing electron. 
Assuming that the incoming electron propagates along the
first axis, we write 
\[
{p'}^i=(p(1-x),\,p'_\bot,0)~,~~~k^i=(px,-p'_\bot,0)\;,
\]
where now $0\leq x\leq 1$ is the fraction of momentum carried away by
the photon. Adopting the ultrarelativistic kinematics to expand 
the energies of the particles, we obtain for the rate of the process,
\[
\begin{split}
\Gamma_{e^-\to\g e^-}
=\frac{\alpha}{2p}\int&\frac{dx d{p'_\bot}^2}{px(1-x)}
\,\delta\bigg(\omega_{LV}'(x)
-\frac{{p'_\bot}^2}{2px(1-x)}\bigg)\\
&\bigg[2\vk p^2(1-x)+\frac{gp^4(1-x)(2-2x+x^2)}{M^2}
+\frac{{p'_\bot}^2}{2(1-x)}\\
&-\frac{\xi p^4}{M^2}x^2(1-x)-\frac{gp^4}{M^2}(1-x)^2
+\frac{g p^4}{M^2}(1-x)(5-5x+x^2)
\bigg]\;,
\end{split}
\]
where
\[
\omega'_{LV}(x)=\vk px-\frac{\xi p^3x^3}{2M^2}
+\frac{gp^3(3x-3x^2+x^3)}{M^2}\;.
\]
The first three terms in the square brackets
come from the LI
part of the matrix element (the first line in (\ref{Mxcher})), while
the rest of the integrand corresponds to the LV part: the second line
in (\ref{Mxcher}). One again observes that the two contributions are
of the same order. Performing the integral over $p'_\bot$ we
obtain 
the differential rate  
\[
\frac{d\Gamma_{e^-\to\g e^-}}{dx}=\alpha\,
\bigg(\frac{2}{x}-2+x\bigg)\,\omega'_{LV}(x)\;. 
\]
To obtain the total rate one has to integrate over the values of $x$
satisfying the condition 
\[
\omega'_{LV}(x)\geq 0\;.
\]
The range of the integration depends on the hierarchies among the LV
parameters. Classifying all possibilities is not relevant for our
discussion. 
In the simplest case $0\leq \vk$, $0\leq g$, $\xi\leq 2g$ 
the integral is over the whole range $0\leq
x\leq 1$ giving
\be
\label{cherratetotal}
\Gamma_{e^-\to \g
  e^-}=\alpha\,p\,\bigg[\frac{4\vk}{3}+\frac{p^2}{M^2}
\bigg(\frac{157g}{60}-\frac{11\xi}{60}\bigg)\bigg]\;.
\ee
A useful characteristic of the process is the rate of the energy loss 
by the
electron 
\be
\label{cherenergy}
\frac{dE}{dt}=-\int dx\; px\,\frac{d\Gamma_{e^-\to\g e^-}}{dx}=
-\alpha\,p^2\bigg[\frac{7\vk}{12}+\frac{p^2}{M^2}
\bigg(\frac{11g}{12}-\frac{2\xi}{15}\bigg)\bigg]\;.
\ee
Taking the ratio of (\ref{cherenergy}) and (\ref{cherratetotal}) 
we observe that on average the emitted photon takes away an order-one
fraction of the electron energy.
In the case $g=\xi=0$
the expressions (\ref{cherratetotal}), (\ref{cherenergy}) reproduce those 
of Ref.~\cite{Klinkhamer:2008ky}.

\subsection{Pair production}
\label{subsec:pair}

We now turn to more complicated reactions containing two particles in
the initial state. The first reaction is production of an electron --
positron pair in the collision of a high-energy photon with a soft
photon from an astrophysical background,
\[
\g\g\to e^+e^-\;.\\
\]
Unlike the two reactions considered in the previous subsections, this process
is kinematically allowed in the LI case. Our goal is to find how its
cross section is affected by LV.

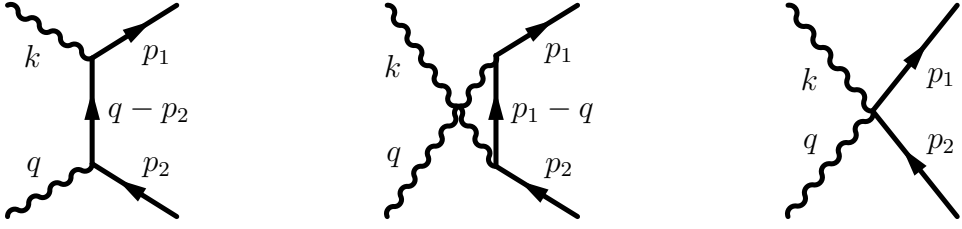
\begin{figure}[h!]
\begin{center}
\begin{fmffile}{pairprod}
\parbox{400pt}
{\begin{fmfgraph*}(80,80)
\fmfpen{thick}
\fmfleftn{i}{2}
\fmfrightn{o}{2}
\fmf{photon,label=$q$}{i1,v1}
\fmf{photon,label=$k$,label.side=right}{i2,v2}
\fmf{fermion,label=$p_2$}{o1,v1}
\fmf{fermion,label=$p_1$}{v2,o2}
\fmf{fermion,label=$q-p_2$,label.side=right}{v1,v2}
\end{fmfgraph*}~~~~~~~~~~~~~~~
\begin{fmfgraph*}(90,80)
\fmfpen{thick}
\fmfleftn{i}{2}
\fmfrightn{o}{2}
\fmf{phantom,tension=0.7}{i1,v1}
\fmf{phantom,tension=0.7}{i2,v2}
\fmf{photon,tension=0.1,label=$q$,label.side=left}{i1,v3}
\fmf{photon,tension=0.1}{v3,v2}
\fmf{photon,tension=0.1,label=$k$,label.side=right}{i2,v4}
\fmf{photon,tension=0.1}{v4,v1}
\fmf{fermion,label=$p_2$}{o1,v1}
\fmf{fermion,label=$p_1$,label.side=right}{v2,o2}
\fmf{fermion,tension=0.7,label=$p_1-q$,label.side=right}{v1,v2}
\end{fmfgraph*}~~~~~~~~~~~~~~~
\begin{fmfgraph*}(80,80)
\fmfpen{thick}
\fmfleftn{i}{2}
\fmfrightn{o}{2}
\fmf{photon,label=$q$,label.side=left}{i1,v}
\fmf{photon,label=$k$,label.side=right}{i2,v}
\fmf{fermion,label=$p_2$,label.side=right}{o1,v}
\fmf{fermion,label=$p_1$,label.side=right}{v,o2}
\end{fmfgraph*}}
\end{fmffile}
\end{center}
\caption{The diagrams contributing to the matrix element
of pair production.
\label{Fig:1}}
\end{figure}

The diagrams contributing to the required matrix element are shown 
in
Fig.~\ref{Fig:1}.  
Note that the third diagram with the two-photon vertex 
is absent in the standard case. 
Denoting the three contributions by ${\cal M}_1$,
${\cal M}_2$, ${\cal M}_3$ we write,
\bseq
\label{Mlin}
\begin{align}
\label{M1lin}
&{\cal M}_1=\bar u(p_1){\cal V}_{1\g}^\m(p_1,p_2-q)\frac{i(\g^\l(\tilde q-\tilde
  p_2)_\l+m)}{(\tilde q-\tilde p_2)_\r(\tilde q-\tilde p_2)^\r-m^2}
{\cal V}_{1\g}^\n(q-p_2,p_2)v(p_2)\varepsilon_\m(k)\varepsilon_\n(q)\;,\\
\label{M2lin}
&{\cal M}_2=\bar u(p_1){\cal V}_{1\g}^\m(p_1,q-p_1)\frac{i(\g^\l(\tilde
  p_1-\tilde q)_\l+m)}{(\tilde p_1-\tilde q)_\r(\tilde p_1-\tilde q)^\r-m^2}
{\cal V}_{1\g}^\n(p_1-q,p_2)v(p_2)\varepsilon_\m(q)\varepsilon_\n(k)\;,\\
\label{M3lin}
&{\cal M}_3=\bar u(p_1){\cal
  V}_{2\g}^{\m\n}(p_1,p_2)v(p_2)\varepsilon_\m(k)\varepsilon_\n(q)\;,
\end{align}
\eseq
where ${\cal V}_{1\g}^\m$ and ${\cal V}_{2\g}^{\m\n}$ 
are given by (\ref{1phot}), (\ref{2phot}) and the four-vectors
with tildes are defined in (\ref{tildep}). We consider the following
kinematical configuration, 
\bseq
\label{momenta}
\begin{align}
\label{phots}
&k^i=(k,0,0)~,&&q^i=(q_x,q_y,0)\;,\\
\label{ferms}
&p_1^i=\bigg(\frac{k+q_x}{2}(1+x),\frac{q_y}{2}(1+x)+p_y,p_z\bigg)\;,
&&p_2^i=\bigg(\frac{k+q_x}{2}(1-x),\frac{q_y}{2}(1-x)-p_y,-p_z\bigg)\;.
\end{align}  
\eseq
The photon with the momentum $q^i$ is assumed to be soft, $q_x,q_y\ll
k$. This allows to 
neglect any LV in the corresponding dispersion relation and write
\[
q^0=\omega\;,~~~ q_x=\omega\cos\theta\;,~~~ q_y=\omega\sin\theta\;,
\]
where $\theta$ is the collision angle ($\theta=\pi$ corresponds to the
head-on collision). 

Derivation of the complete expression 
for the square of the matrix element in this case would be too
cumbersome and not illuminating. Instead, our strategy will be to
compute the matrix element in the leading approximation expanding 
in the small
ratio $\omega/k$. To this end we need to determine the orders of
magnitude of various quantities appearing in the calculation.
The phenomenologically interesting case is when
the LV corrections in the particle dispersion relations (\ref{Eg}),
(\ref{Ee}) 
are of the same order as the square of 
the relativistic invariant mass $s=2k\omega(1-\cos\theta)$. 
In other words, we shall
treat the LV corrections $\vk k^2$, $g k^4/M^2$,
$\xi k^4/M^2$ as being of the same order as $k\omega$.
Besides, in the standard LI case the perpendicular components of the
electron and positron momenta are also determined by the invariant
mass,
\[
p_y^2,~ p_z^2\sim k\omega\;.
\]
We will assume this estimate to hold in the presence of LV, as will be
verified by the explicit calculation below.

We now observe that the denominators of 
the propagators in the
amplitudes (\ref{M1lin}), (\ref{M2lin}) are of order $k\omega$,
\[
(\tilde q-\tilde p_2)_\l(\tilde q-\tilde p_2)^\l-m^2\approx
-k(\omega-q_x) (1-x)~,~~~~
(\tilde p_1-\tilde q)_\l(\tilde p_1-\tilde q)^\l-m^2\approx
-k(\omega-q_x) (1+x)\;,
\]
where we have used
\[
\tilde {p_1}_\l\tilde p_1^\l=\tilde {p_2}_\l\tilde p_2^\l=m^2\;.
\]
Therefore to get the leading-order result we need to calculate the
numerator in the expression for $\overline{|{\cal M}_1+{\cal M}_2|^2}$
to order $O\big((\omega/k)^2\big)$. 
On the other hand, ${\cal M}_3$ is
already suppressed by the first power of the LV parameters. Thus we
can neglect its square while in the interference terms 
$\overline{{\cal M}_3{\cal M}_1^*}$, $\overline{{\cal M}_3{\cal M}_2^*}$
it is
enough to take only the LI part of ${\cal M}_1$, ${\cal M}_2$.
After a rather tedious but straightforward calculation we obtain
\be
\label{pairMatthet}
\begin{split}
\overline{|{\cal M}|^2}=e^4\bigg[4\frac{1+x^2}{1-x^2}
-\frac{32p_\bot^2}{k(\omega-q_x)(1-x^2)^2}
+\frac{64 p_\bot^4}{k^2(\omega-q_x)^2(1-x^2)^3}
-\frac{16\omega_{LV}(x) p_\bot^2}{k(\omega-q_x)^2(1-x^2)}
\bigg]\;,
\end{split}
\ee
where $p_\bot^2=p_y^2+p_z^2$ and $\omega_{LV}(x)$ is defined in (\ref{omegaLV}). 
In deriving (\ref{pairMatthet}) 
we have neglected the electron mass, which is justified well
above the threshold\footnote{We will need to take the
  mass into account later in order to cut off the logarithmic
  divergence in
the phase space integral.}. 
It is worth mentioning that in the calculation leading to (\ref{pairMatthet})
one finds a large amount of cancellations between the contributions
from various
products of the amplitudes (\ref{Mlin}). Thus all
terms quadratic in the LV parameters $\vk$, $g$, $\xi$
disappear. Besides,  
the two interference terms containing ${\cal M}_3$ cancel each other. 
The latter is
an artifact of the massless approximation: one can check that the
contribution of the
two-photon vertex does not vanish if the
finite electron mass is taken into account. 

The cross section is given by the formula\footnote{The combination
  $(1-\cos\theta)$ in the denominator of the prefactor 
comes from the projection of the relative velocity of the colliding
photons on the $x$-axis, 
$|v_{k,x}-v_{q,x}|$. This combination enters in the relativistic
definition of the invariant cross section. Though in
our case the relativistic invariance is absent, we prefer to stick to
the textbook definition to facilitate comparison with the standard
results. Note that in the LI case the above prefactor
  turns into relativistic invariant inversly proportional to the
  square of the invariant mass~$s$.} 
\be
\label{crossggthet}
\sigma_{\g\g\to e^+e^-}=\frac{1}{32\pi k\omega(1-\cos\theta)}
\int dx\frac{dp_\bot^2}{k(1-x^2)}\;\delta\bigg(\omega-q_x+\omega_{LV}(x)-
\frac{2p_\bot^2}{k(1-x^2)}\bigg)\overline{|{\cal M}|^2}\;.
\ee
 Substituting $q_x=\omega\cos\theta$ and integrating 
over $p_\bot$ we obtain
\[
\sigma_{\g\g\to e^+e^-}=
\frac{\alpha^2\,\pi}{2 k\omega(1-\cos\theta)}
\int dx\,\frac{1+x^2}{1-x^2}\;\bigg[1
+\bigg(1+\frac{2\omega_{LV}(x)}{\omega(1-\cos\theta)}\bigg)^2\bigg]\;.
\]
The domain of integration in $x$ is determined by the condition,
\be
\label{domthet}
\omega(1-\cos\theta)+\omega_{LV}(x)>0\;.
\ee
In what follows we restrict to 
the case when it covers the whole range $-1\leq
x\leq 1$. Then the integral logarithmically diverges at the
end-points. These correspond to the kinematical configuration when the
total energy is carried away by one of the fermions, while the second
fermion stays with zero energy. Clearly, such configuration is
possible only for strictly massless fermions and the divergence is
cut off once we take into account the finite electron mass. 
This is
achieved by substituting 
\be
\label{masstrick}
p_\bot^2\mapsto p_\bot^2+m^2
\ee   
in the argument of the $\delta$-function in (\ref{crossggthet}). 
Consequently the condition (\ref{domthet}) is replaced by 
\[
p_\bot^2>0\Longrightarrow 1-x^2>\frac{2m^2}{k(\omega(1-\cos\theta)
+\omega_{LV})}\;,
\]
implying that $|x|$ is strictly less than one.
This yields the total cross section with the logarithmic accuracy,
\be
\label{crossggthet2}
\sigma_{\g\g\to e^+e^-}=
\frac{\alpha^2\,\pi}{ k\omega(1-\cos\theta)}
\bigg[1
+\bigg(1+\frac{2\omega_{LV}}{\omega(1-\cos\theta)}\bigg)^2\bigg]
\log\bigg[\frac{k(\omega(1-\cos\theta)+\omega_{LV})}{m^2}\bigg]\;,
\ee
where $\omega_{LV}$ is taken at $x=1$.
This formula is valid as long as 
\[
k(\omega(1-\cos\theta)+\omega_{LV})\gg m^2\;.
\]
We see that the effect of LV on the cross section is 
governed by the ratio 
$$r=\frac{\omega_{LV}}{\omega(1-\cos\theta)}\;.$$ 
The condition (\ref{domthet}) implies $r>-1$. 

One observes that the 
cross section can be significantly enhanced if $r$ is larger than
one. However, in this case $\omega_{LV}$ is positive
implying that the photon decay is kinematically allowed. The
latter will dominate the pair production in most astrophysical
circumstances. From this viewpoint, the case of $r$ belonging to the
interval $-1<r<0$ is more interesting.
Then the photon decay is forbidden, 
but the pair production
on the background still takes place. In this case the cross section
(\ref{crossggthet2}) differs from the standard relativistic expression
by a factor of order one.

\subsection{Pair production in the Coulomb field}
\label{subsec:coulomb}

The last reaction we consider is 
pair production by a high-energy photon in the Coulomb field of a
nucleus,
\[
\g Z\to Z e^+e^-\;.
\]
Here $Z$ denotes the charge of the nucleus in the units of the
elementary charge. This reaction does not have a threshold and
in the standard LI case dominates the interaction of the UHECR
photons with the Earth atmosphere giving rise to the
electromagnetic showers that are used for the detection and
identification of such photons \cite{Risse:2007sd}. 
The analysis of the changes induced by LV in the cross section of this
reaction is therefore crucial to determine the detection efficiency
for UHECR photons in LV models.

The process is conveniently represented as a collision of the
high-energy photon with a soft virtual photon from the nucleus' Coulomb
field. Thus it is 
described by the same diagrams shown in
Fig.~\ref{Fig:1}, as the two-photon collision of the previous
subsection. 
For the momenta of the particles taking part in the reaction we use
the parameterization (\ref{momenta}) and evaluate the square of the
matrix element to the leading order in the small quantity $q_x/k$. The
LV contributions appearing in the dispersion relations (\ref{Eg}),
(\ref{Ee}) will be assumed to be of order~$kq_x$.

There are several simplifications compared to the case of the previous
subsection. First, the virtual photon has purely time-like polarization,
$\varepsilon_\m(q)=\delta_\m^0$, 
and thus the contribution of the third diagram in
Fig.~\ref{Fig:1} vanishes identically, see (\ref{2phot}). Second, it
has vanishing energy, $q^0=0$, which reduces the number of terms
appearing in the calculations. Moreover, it turns out that the leading
contribution in the numerator of the square of the matrix element is
of order 
$O(kq_x)$ (as opposed to
$O\big((kq_x)^2\big)$ 
in the case of collision of two
real photons). This means that it is sufficient to evaluate the
numerator up to linear approximation in the LV parameters.

On the other hand, unlike the case of the previous subsection, one
cannot assume the components $q_x$, $q_y$ of the virtual photon
momentum to be of the same order, as the calculation of the cross
section involves integration over all possible values. Instead, we are
going to find that the integral is saturated at
\be
\label{qyqx}
q^2_y\sim kq_x~~\Longleftrightarrow~~ q_y\gg q_x\;.
\ee
Thus we have to keep all terms with $q_y$ up to second power.

Finally, in the matrix element we will
provisionally neglect the electron mass. The conditions for the 
validity of this
approximation will be discussed below. Then the direct computation
yields,
\be
\label{Mnucl}
\begin{split}
\overline{|{\cal M}|^2}=\frac{2Z^2e^6k^2}{(q_x^2+q_y^2)^2}\bigg[
&\frac{1-x}{1+x}\cdot
\frac{4p_y^2+4p_z^2+4p_yq_y(1+x)+q_y^2(1+x)^2-k\omega_{LV}(x)(1-x^2)^2}
{(2p_yq_y+q_y^2x-kq_x(1-x))^2}\\
+&\frac{1+x}{1-x}\cdot
\frac{4p_y^2+4p_z^2-4p_yq_y(1-x)+q_y^2(1-x)^2-k\omega_{LV}(x)(1-x^2)^2}
{(2p_yq_y+q_y^2x+kq_x(1+x))^2}\\
+&2\frac{4p_y^2+4p_z^2+4p_yq_yx-k\omega_{LV}(x)(1-x^2)^2}
{(2p_yq_y+q_y^2x-kq_x(1-x))(2p_yq_y+q_y^2x+kq_x(1+x))}\bigg]\;.
\end{split}
\ee
Note the factor
\be
\label{Coul}
\frac{Z^2 e^2}{(q_x^2+q_y^2)^2}
\ee
describing the density of virtual photons in the Coulomb field. 
In deriving (\ref{Mnucl})
we have summed over the spins of the electron and positron and
averaged over the polarizations of the high-energy photon. 

The formula for the cross section in the case of scattering on a fixed
scattering center reads,
\[
\sigma_{\g Z\to Ze^+e^-}=\frac{1}{2k}\int\frac{d^3 p_1}{(2\pi)^3 2E_1}
\frac{d^3 p_2}{(2\pi)^3 2E_2}\, (2\pi)\, \delta(E_\g-E_1-E_2)\,
\overline{|{\cal M}|^2}\;.
\]
It is convenient to trade the integration variable $p_2^i$ for
$q^i=(k-p_1-p_2)^i$. Then, using the axial symmetry of the problem, we
can perform integration over the direction of $q^i$ in the
$yz$-plane. This gives a factor $2\pi$ and leaves us with the
integrals over $q_x$ and $q_y$. The first is removed with the help of
the $\delta$-function using the expansion
\be
\label{Eexpand}
E_\g-E_1-E_2\approx \omega_{LV}(x)-\frac{2(p_y^2+p_z^2)}{k(1-x^2)}
-\frac{q_y^2}{2k}-q_x\;.
\ee
Notice that this restricts the integral to the portion of the phase
space where
\[
q_x\sim |\omega_{LV}|~,~~~~q_y^2,\, p_y^2,\, p_z^2\sim k|\omega_{LV}|\;.
\]
In particular, (\ref{qyqx}) is indeed fulfilled. 
Neglecting $q_x$ in the Coulomb propagator\footnote{This approximation
  breaks down in a certain corner of the phase space, see
  below.} (\ref{Coul}) and
splitting the integral over $p_1^i$ into the
longitudinal and transverse parts we
arrive at
\be
\label{sigZ1}
\begin{split}
\sigma_{\g Z\to Ze^+e^-}
=\int dx\, dp_y\,dp_z\,dq_y\,\frac{2Z^2e^6}{(2\pi)^4q_y^3}
\bigg[
&\frac{4p_z^2+(2p_y+q_y(1+x))^2-k\omega_{LV}(x)(1-x^2)^2}
{D_1^2}\\
+&\frac{4p_z^2+(2p_y-q_y(1-x))^2-k\omega_{LV}(x)(1-x^2)^2}
{D_2^2}\\
-&\frac{2(4p_y^2+4p_z^2+4p_yq_yx-k\omega_{LV}(x)(1-x^2)^2)}
{D_1 D_2}\bigg]\;,
\end{split}
\ee
where
\bseq
\label{DDs}
\begin{align}
D_1=4p_z^2+(2p_y+q_y(1+x))^2-2k\omega_{LV}(x)(1-x^2)\;,\\
D_2=4p_z^2+(2p_y-q_y(1-x))^2-2k\omega_{LV}(x)(1-x^2)\;.
\end{align}
\eseq
In what follows we will restrict to the case $\omega_{LV}(x)<0$, so
that the denominators (\ref{DDs}) never
vanish. Physically, this corresponds to the parameter region where
spontaneous photon decay is forbidden.

The next step is to integrate over $p_y$, $p_z$. Note that the
contribution of each
term in the square brackets in (\ref{sigZ1}), when considered
separately, 
is logarithmically
divergent at \mbox{$p_y,p_z\to\infty$}. However, these
divergences cancel in the sum. We obtain,
\[
\sigma_{\g Z\to Ze^+e^-}\!\!=\alpha^3Z^2\!\!
\int\! dx\;\frac{1+x^2}{k|\omega_{LV}(x)|}
\int\!\frac{dy}{y^2}\bigg[\frac{y+1-x^2}{\sqrt{y(y+2(1-x^2))}}
\log\bigg[\frac{\sqrt{y+2(1-x^2)}+\sqrt{y}}
{\sqrt{y+2(1-x^2)}-\sqrt{y}}\bigg]-1\bigg]\,,
\]
where 
\[
y=\frac{q_y^2}{k|\omega_{LV}(x)|}\;.
\]
The $y$-integral logarithmically diverges at $y\to 0$. 
This is an artifact of neglecting 
$q_x$ in the Coulomb propagator. 
Given that $q_x$ is of order $|\omega_{LV}|$, we conclude that
the $y$-integral must be cut off at 
\be
\label{ynoscreening}
y_0\sim \frac{|\omega_{LV}|}{k}\;.
\ee
This leads to the expression
\be
\label{sigZ3}
\sigma_{\g Z\to Ze^+e^-}=\frac{2Z^2\alpha^3}{3}\int\frac{dx}{k|\omega_{LV}(x)|}\,
\frac{1+x^2}{1-x^2}\,\bigg[\log\frac{1-x^2}{y_0}+\frac{13-6\log2}{6}\bigg]\;.
\ee
Here we have expanded the integrand under the assumption
\be
\label{y0small}
y_0\ll (1-x^2)\;,
\ee
whose validity will be checked shortly. 
The above integral is again logarithmically divergent at 
$x\to\pm 1$. As in the previous subsection, to cut off this
divergence we have to take into account the finite 
electron mass. This is achieved by the substitution 
(\ref{masstrick}) in the argument of the energy-conservation
$\delta$-function. From (\ref{Eexpand}) we find 
that the mass can be neglected as long as
\be
\label{xxs}
1-x^2\gg\frac{m^2}{|k\omega_{LV}(x)|}\;.
\ee   
This is satisfied in most of the integration domain
provided 
the hierarchy
\be
\label{hierarchy}
|k\omega_{LV}|\gg m^2\;,
\ee
which represents the condition for the validity of our
approximation. 

Restricting the domain of integration in (\ref{sigZ3}) according to 
(\ref{xxs}) we obtain the total cross section with 
logarithmic accuracy,
\be
\label{sigZ4}
\sigma_{\g Z\to Ze^+e^-}=\frac{4Z^2\alpha^3}{3k|\omega_{LV}|}
\bigg[\log\frac{k}{|\omega_{LV}|}
-\frac{1}{2}\log{\frac{k|\omega_{LV}|}{m^2}}\bigg]
\log\frac{k|\omega_{LV}|}{m^2}\;,
\ee
where $\omega_{LV}$ is taken at $x=1$.

In the realistic situation the nucleus is surrounded by atomic
electrons that screen its Coulomb field at large distances. Thus $q_y$ is
bounded from below by the inverse size of the atom, $q_y\gtrsim
1/a$. In the
mean-field atomic model one finds (see e.g. \cite{LL}),
\[
a\sim\frac{1}{\alpha Z^{1/3}m}\;.
\]  
If the momentum
$1/a$ is larger than $|\omega_{LV}|$, we have to replace 
(\ref{ynoscreening}) by 
\be
\label{yscreening}
y_0\sim\alpha^2 Z^{2/3}\frac{m^2}{|k\omega_{LV}|}\;.
\ee
The rest of the analysis goes as before
and yields
\be
\label{sigZ5}
\sigma_{\g Z\to Ze^+e^-}=\frac{4Z^2\alpha^3}{3k|\omega_{LV}|}
\bigg[2\log{\frac{1}{\alpha Z^{1/3}}}
+\frac{1}{2}\log\frac{k|\omega_{LV}|}{m^2}\bigg]
\log\frac{k|\omega_{LV}|}{m^2}\;.
\ee
It remains to check the assumption (\ref{y0small}). Comparing
(\ref{xxs}) with (\ref{ynoscreening}), (\ref{yscreening}) we find that
it is equivalent to the requirement $|\omega_{LV}|\ll m$ (in the case
of no screening), or $\alpha Z^{1/3}\ll 1$ (with screening). These
conditions are satisfied for real nuclei and phenomenologically
interesting values of the LV parameters. 

The expressions (\ref{sigZ4}), (\ref{sigZ5}) must be compared to the
standard result 
\cite{Bethe:1934za} 
\[
\sigma_{\g Z\to Ze^+e^-}^{LI}=\frac{28Z^2\alpha^3}{9 m^2}\times
\begin{cases}
\log{\frac{2k}{m}} - \frac{109}{42}& ~~~~\text{no screening}\\
\log{\frac{183}{Z^{1/3}}}-\frac{1}{42} &~~~~\text{with screening}
\end{cases}
\]
We see that in the regime (\ref{hierarchy}) LV strongly
suppresses the cross section of pair production on nuclei. 
Besides, the LV cross section is dominated, 
according to (\ref{sigZ3}), by the configurations when one of the
produced fermions carries
most of the energy ($|x|\approx
1$). This is in contrast to the standard QED where the energy
distribution of the pair is smooth over the whole range $-1\leq x\leq 1$. 
Physical
consequences of the above effects for registration of UHECR photons will be
reported elsewhere \cite{RSS}.
 
\section{Discussion}
\label{sec:discussion}

We have systematically derived the Feynman rules for a model of LV QED
and have applied them to calculate the rates of several
astrophysically relevant processes. Our analysis demonstrates that to
find the precise result one must take into account both kinematical
and dynamical aspects of LV whose effects on the rates are of the same
order. The first --- kinematical --- class of effects amounts to the
change of the phase-space integrals due to modified dispersion
relations of the particles. While the dynamical effects include the
modifications of the matrix elements due to the changes in the
particles' wavefunctions and in the interactions vertices. It is worth
stressing that the gauge invariance relates the structure of the
interaction vertices to the dispersion relations. Thus the
modification of the vertices is unavoidable in gauge theories with LV.

The Feynman rules formulated in this paper can be straightforwardly
applied to calculation of other cross sections in QED with LV. An
interesting process from the astrophysical viewpoint is the photon
splitting 
$\g\to 3\g$
that becomes kinematically 
allowed whenever the parameter $\xi$ in the photon 
dispersion relation (\ref{Eg}) is positive. However, technically the
calculation of the corresponding rate appears very
challenging. Indeed, the process goes via the fermion loop and apart
from the standard box diagram will involve the graphs with the
insertions of the multi-photon vertices 
(\ref{2phot}), (\ref{3phot}). 
This drastically increases the number of topologically distinct
contributions. The issues related to the treatment of the divergences
appearing in the loop integrals further complicate the task.

A technically more promising direction is extension of the approach
put forward in this paper to include the electroweak and hadronic
processes. Of particular interest is an accurate analysis of the LV
effects on the rates of various process involving cosmogenic neutrinos. 

As the concluding remark we note that 
the results of this paper provide justification
of a heuristic method often used in the literature to make
order-of-magnitude estimates of LV effects \cite{Coleman:1998ti}. The
reasoning goes as follows. One observes from the dispersion relations
(\ref{Eg}), (\ref{Ee}) that LV introduces effective momentum-dependent
masses for the photon and electron,
\begin{align}
&m_\g^2(k)\equiv E_\g^2-k^2=\frac{\xi k^4}{M^2}\;,\notag\\
&m_e^2(p)\equiv E_e^2-p^2=m^2+2\vk p^2+\frac{2g p^4}{M^2}\;.\notag
\end{align}
These masses set the scale of the characteristic energy-momentum 
transfer in
various processes and can be used to estimate the corresponding cross
sections on dimensional grounds. For example, consider the photon
decay. Treating the photon as
massive, let us perform a boost into its rest
frame\footnote{Of course, this reasoning applies only to the case
  $m_\g^2>0$}. In that frame the energy released in the photon decay
is of the order $m_\g$, which gives for the width in this frame
\[
\Gamma_{\g\to e^+e^-}^{\mathrm{rest}}\sim\alpha\, m_\g\;.
\]
To obtain the width in the original frame, we have to perform the
reverse Lorentz transformation. This introduces a
time-dilation factor $m_\g/k$. In this way one obtains 
\[
\Gamma_{\g\to e^+e^-}\sim \alpha\,m_\g^2(k)/k\;.
\]
The exact formula (\ref{gdectot}) is more complicated and depends also
on $m^2_e$ in a
non-trivial way. However, for $m_e\sim m_\g$ the simple derivation
outlined above gives the correct oder-of-magnitude estimate.
Similarly, for pair production in the Coulomb field
the momentum transfer between the nucleus and the photon is of order
$2m_e(k)$. This gives on dimensional grounds,
\[
\sigma_{\g Z\to Ze^+e^-}\sim \frac{Z^2\alpha^3}{m_e^2(k)}\;,
\]
where we have included the appropriate powers of the fine-structure
constant and the nucleus' charge.
For $m_\g^2\lesssim m_e^2$ this estimate coincides with the
exact expressions (\ref{sigZ4}), (\ref{sigZ5}) up to the logarithmic
factors.

\paragraph*{Acknowledgements}

We are grateful to Sergei Demidov,
Dmitry Gorbunov, Stefano Li\-be\-ra\-ti, Maxim Libanov, 
Emin Nugaev, Alexander Panin,
Valery Rubakov and Sergey Troitsky for useful
discussions. We thank Fedor Bezrukov for encouraging interest. 
This work was supported in part by 
the Grants of the President of Russian Federation
NS-5590.2012.2, MK-1632.2011.2 (G.R. and P.S.) 
and MK-3344.2011.2~(P.S. and S.S.), the RFBR grants
10-02-01406 (G.R.), 11-02-92108 (S.S.),
 11-02-01528 (G.R. and S.S.), 12-02-01203 (S.S.), 12-02-91323 (G.R.)
and by the Dynasty Foundation (P.S. and S.S.).

\end{document}